# A wireless secure key distribution system with no couriers: a One-Time-Pad Revival

Geraldo A. Barbosa

*Abstract*— Among the problems to guarantee secrecy for in-transit information, the difficulties involved in renewing cryptographic keys in a secure way using couriers, the perfect secrecy encryption method known as One-Time-Pad (OTP) became almost obsolete. Pure quantum key distribution (QKD) ideally offers security for key distribution and could revive OTP. However, special networks that may need optical fibers, satellite, relay stations, expensive detection equipment compared with telecom technology and the slow protocol offer powerful obstacles for widespread use of QKD. Classical encryption methods flood the secure communication landscape. Many of them rely its security on historical difficulties such as factoring of large numbers -- their alleged security sometimes are presented as the difficulty to brake encryption by brute force. The possibility for a mathematical breakthrough that could make factoring trivial are poorly discussed. This work proposes a solution to bring perfect secrecy to in-transit communication and without the above problems. It shows the key distribution scheme (nicknamed KeyBITS Platform) based on classical signals carrying information but that carry with them recordings of quantum noise. Legitimate users start with a shared information of the coding bases used that gives them an *information advantage* that allows easy signal recovery. The recorded noise protects the legitimate users and block the attacker's access. This shared information is refreshed at the end of each batch of keys sent providing the secret shared information for the next round. With encryption keys distilled from securely transmitted signals at each round OTP can be revived and at fast speeds.

## I. Introduction

INFORMATION is a valued commodity in both military and non-military enterprises. It loses its value when it is unprotected and usable by competitors and those who would inflict harm. Perfect secrecy is required to safeguard information at top secret level.

**Perfect secrecy** and **One-time pad encryption** - Shannon [1] defined "Perfect Secrecy" in cryptography: "*Perfect Secrecy is defined by requiring of a system that after a cryptogram is intercepted by the enemy the a posteriori probabilities of this cryptogram representing various messages be identically the same as the a priori probabilities of the same messages before the interception. It is shown that perfect secrecy is possible but requires, if the number of messages is finite, the same number of possible keys*". The number of key bits equal to the number of bits in the message is, in essence, Vernam's idea for encryption (U.S. Patent 1,310,719. - US, 1919), that later was perfected by Maugborne [2] by adding random information in a punched paper tape. This started a practical form for the **one-time pad** (OTP) encryption. OTP cannot be broken even by an ideal quantum computer – it is a very precious asset not to be forgotten.

Text can be represented in computers and telecommunication by the so called ASCII codes where any character can be mapped, for example, to a binary sequence (0s or 1s). In binary form one-time-pad encryption amounts to the bitwise XOR (Exclusive OR) operation over two bits B and K that obeys the Truth Table I where B may designate a bit in a byte representing a digitized letter in a message and K the encryption key bit (0 or 1).

TABLE I
*XOR Truth Table*

| XOR Truth Table | |  |
|---|---|---|
| Input | | Output |
| B | K | B **XOR** K |
| 0 | 0 | 0 |
| 0 | 1 | 1 |
| 1 | 0 | 1 |
| 1 | 1 | 0 |

Recovery of a message $B$, OTP encrypted with key $K$ (= $B$ XOR $K$), is straightforward: just XOR it with $K$. It results ($B$ XOR $K$)XOR $K = B$.

As shown in Appendix XI.A one can *measure* the amount of "perfect secrecy" by calculating the Mutual Information $I(X;Y)$ between variables $X$ and $Y$. $I(X;Y)$ measures the amount of $X$ if $Y$ is known. $X$ could be, say, the plaintext message $M$ and $Y$ the encrypted form $C$ accessed by the attacker. In particular, for *key distribution* (the case of our interest), the plaintext will be just a sequence of random bits, with no pattern associated to it; and encryption will be applied to this bit sequence.

The calculation assuming encryption by one-time-pad gives $I(M;C) = 0$, what assures a complete failure by the attacker in obtaining $M$ - even with the attacker having a perfect copy of the transmitted signals. It should be emphasized that standard

G. A. Barbosa, Full Professor, Universidade Federal de Minas Gerais / Brazil (up to 1995) / Northwestern University (2000-2012). Pres. – KeyBITS Encryption Technologies LLC/MD-US. Chief Scientist – Diana Rae Carl LLC/VA-US. (e-mail: geraldoabarbosa@gmail.com).

telecommunication signals are being used: they can be precisely copied by the attacker

**Encryption nowadays** - This perfect secrecy level result using classical signals is not replicated by many of the classical encryption methods currently offered. For example, public-key encryption and digital signatures rely on the historically-hard mathematical problems of factoring. Quite unfortunately there are *no* proofs to guarantee their secrecy level - instead they look for computational difficulties to break keys by brute force and accept that, with the use of long keys for RSA or algorithms based on elliptic curves, security will result. Basically, security has relied on unproven hypotheses and unsupported by existing physical and mathematical tenets. *Faith is not enough*.

Additional menaces to these unproven methods are Quantum Computation possibilities. Better encryption methods have to be found. Some of the current options involving support from physics are these:

**QKD** - The pure Quantum Key Distribution (QKD) methods can guarantee secrecy in key distribution and, therefore, would allow application of one-time-pad encryption (=symmetric encryption) with the distilled keys. However, the need for a quantum communication dedicated channel, the slow method and the overall high costs associated with QKD give no indication for adoption of QKD for wide use in a near future.

**$M$-ry encoding and physical noise in optical channels** - The key distribution method encoding bits in $M$-ry bases with superposed quantum noise is a variation for key distribution that can bring Perfect Security and it was developed for optical channels, where the quantum noise is inherent to the optical channel itself [3]. Although the system is lower cost than QKD, it is faster and has a longer distance range, it needs optical channels such as optical fiber networks. The literature is relatively vast (see examples from [3] to [25]).

One could ask if a different combination of methods could create systems allowing proved and *fast secure* communication at *low cost* and working in a *generic channel*.

**Wireless $M$-ry coding with added physical noise in classical channels** - This work meets the above criteria, it uses classical signals in standard communication channels to distribute keys in a secure way. The achieved security is made possible by the use of quantum fluctuations of optical origin that were *recorded* and *added* to the bit signals coded in $M$-ry levels (explained ahead). For each $M$-ry coded bit sent a new recorded noise signal is added that cloaks the $M$-ry level used and does not allow the attacker to obtain the bit sent. This system, known as the KeyBITS Platform, will be presented and discussed in detail in the next sections. As the one-time-pad encryption with keys unknown to the attacker gives perfect secrecy, it should be concluded that the perfect secrecy in encryption has to be provided by the secure transmission of the encryption keys by the KeyBITS Platform.

The security of this transmission starts with the fact that optical quantum noise is irreducible (cannot be reduced or eliminated) in principle, regardless any technological capability of the adversary.

Furthermore, the process of recording *instances* of this quantum noise (a classical procedure) and adding it to each coded bit sent is also irreducible for the adversary; nothing in the transmission channel can be done to diminish or eliminate it.

In the wireless case or in an optical channel a measurement by an adversary always include noise. In the wireless case this recording is done within the KeyBITS Platform and transmitted to the public channel. This is the *fundamental* protection used in the KeyBITS technology.

As will be shown, a classical privacy amplification process will also be applied with the function of discarding any infinitesimally information leak to the adversary. It also avoids attacks on past or future keys in case the adversary succeeds in obtaining a sequence of used keys from a transmission round.

For authentication purposes the transmitted signals can be subject to an additional layer of encryption based on conventional public key cryptography (PK) allowing compliance with certification standards. Zeroing of all the critical security information can be forced upon any tamper trial over the Platform. The KeyBITS Platform can be set to satisfy the most stringent levels of the Security Requirements for Cryptographic Modules (FIPS 140-2).

The first part of the paper provides a general description and a few technical details about the KeyBITS technology. More details are given in the Appendix. The Appendix also has a section comparing the KeyBITS Physical Random Bit Generator with other physical generators. A final section compares rough order of magnitude costs between QKD and KeyBITS. Finally, steps required to achieve a fully functional KeyBITS Platform are posited

## II. KEYBITS PLATFORM

THE KeyBITS Platform is designed to *generate* cryptographic keys, to *distribute* them through classical channels in a *secure* way, and to do so *without using couriers*. These functions will be discussed in this paper. First, the generation of bits is discussed and second, the secure distribution process is explained relative to its dependence on physical noise and on a privacy amplification process.

### A. *Entropy source for the Physical Random Bit Generator*

The generation of keys uses optical quantum fluctuations in a laser beam. A light field, similar to that of an amplitude stabilized laser, also known as a coherent field, presents spontaneous fluctuations of photon numbers of an uncontrollable character. These fluctuations of quantum origin are also known as "optical shot noise".

Using a laser with a long coherence time $\tau$ (time where the laser keeps a constant phase $\phi$) and taking light intensity samples within short time windows of duration $\Delta t$ ($\Delta t \ll \tau$), fluctuations can be seen. They are directly related to the photon number fluctuations $\langle (n - \langle n \rangle)^2 \rangle \equiv \sigma_n^2$ characteristics of the coherent field.

The light intensity presents an average value $\langle I \rangle$ around which the fluctuations $\Delta I$ (or $\Delta n$) occur. Fluctuations occurring above or below the average physical signals are recorded and represent the desired bits. These signal generations cannot be reproduced by any algorithm and are purely of a quantum origin. This entropy source produces bits in a completely different process from pseudo-random generators that utilize algorithms to produce random-like sequences of bits but have a deterministic characteristic at their core.

Figure 1 sketches the elements that constitute the Physical Random Bit Generator (PhRBG), which generates keys from quantum fluctuations in a light field. Details can be seen in Appendix X.B.

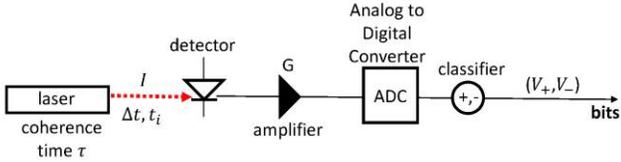

Fig. 1. A light beam is intensity sampled by a fast detector at instants $t_i$ within time windows $\Delta t$. The resulting current is amplified (G). The analog signals pass to an Analog to Digital Converter (ADC) and the digitized voltage levels are classified above or below average producing a stream of random voltages $(V_+, V_-)$ representing the physical bits.

Figure 2 shows the PhRBG. Records of the quantum fluctuations (optical shot noise) are also shown in the computer display. Comparisons of the keyBITS PhRBG with other physical random number generators can be seen in Appendix X.L

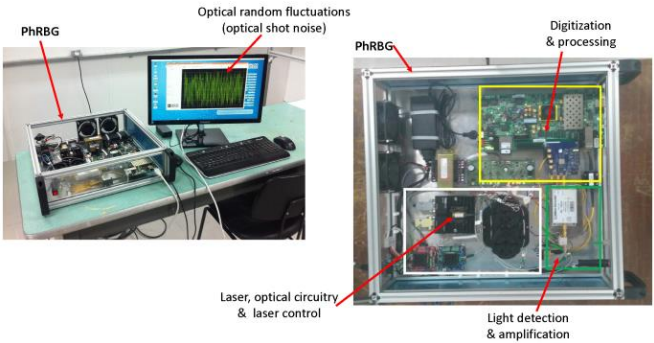

Fig. 2. On the left is a PhRBG on a table. Note that the display shows the random fluctuations obtained from the laser beam. On the right are components of the PhRBG.

### B. Key distribution: Transmission and Receiving Stations

This distribution process is an evolution over a similar scheme carried over an optical channel, where the cloak effect produced by the $M$-ry coded signals was imposed by the optical fluctuations existing in the optical carrier itself, the laser beam. Now the same cloak effect is achieved by adding over every classical coded bit signal a distinct recorded noise component. These noisy signals are recorded *instances* of measurements (samplings) taken on the quantum fluctuation signals. While in the optical channel the noise is always present, in the classical channel it has to be added bit-by-bit.

Random keys generated by the PhRBG in a emitter station A are coded (discussed below), transmitted and received by $N$ receiving stations $(1,2,3,...N)$. See Figure 3. Station A contains the PhRBG, an optical noise source, light detectors, amplifiers and processing electronics and software; it is hardware based with dedicated software. These elements are the main components of the KeyBITS Platform. The Platform is kept within an air gap (=region with no continuous direct contact between the interior and the exterior) to avoid direct attacks by hackers. Communication with the outside is controlled to minimize attacks. Signals coming out from the Platform inside the air gap reach a PC or processor with an outside IP connection or any other private radio network that will direct the generated keys to the $N$ receiving stations.

The receiving stations have a receiving unit inside an air gap and a PC or processor with the software to process the received information. Sometimes the letters A and B are used for these stations.

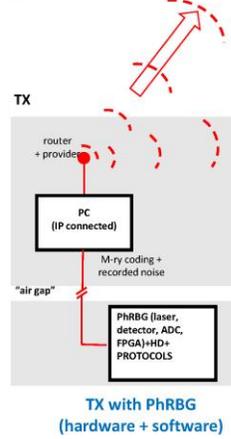 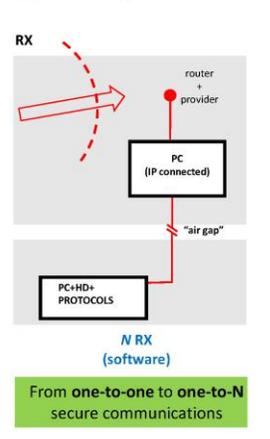

Fig. 3. At left the key transmission station (TX) is shown where the KeyBITS Platform is inside an air gap. Communication with the receiving stations (RX) is done by a PC or processor that has a controlled communication with the Platform. The RX stations are software-based and communication with the exterior to the processing PC is accomplished in a controlled way. Signals sent from TX to the RX are processed in a similar way inside TX and inside any RX so that the same distilled keys result in TX and RXs. The processing unit for key distillation is called "Bit Pool". RX stations will decode the sent signals to extract the signal bits originally generated by the PhRBG in TX.

The receiving stations are mainly software-based and with capacity to process the received raw key stream. This processing has the purpose of obtaining a distilled fresh key stream to be used for encryption. The same distillation process in done in the emitting station so that the emitter and the $N$ receivers become equipped with identical sets of distilled keys necessary for symmetric encryption.

Due to the acceptance of general communication channels the system can choose a different network route in case of rupture (forced or accidental) of the channel in use. This may provide a very resilient system against disruptive attacks to the communication channels available to the users.

### C. KeyBITS Platform blocks

The KeyBITS Platform can be roughly represented as a three blocks with distinct functionalities: The PhRBG, the Noise Generator and the Bit Pool. See Figure 4.

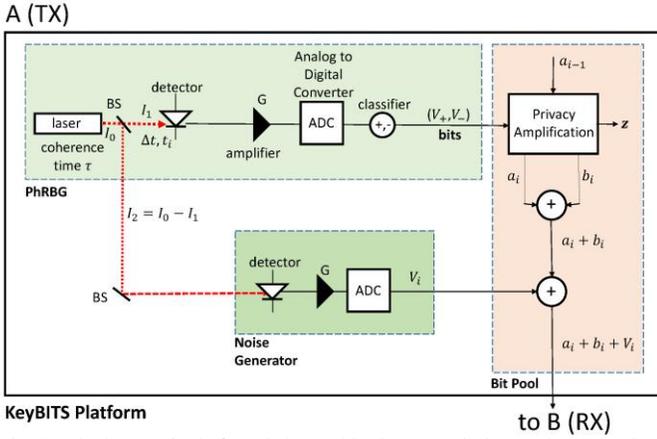

Fig. 4. The keyBITS Platform is located in the transmission station TX. Three blocks compose the Platform: The PhRBG, a Noise Generator, and a Bit Pool. The PhRBG is the key generator modulus where keys are generated from quantum fluctuations in a laser beam. They are detected, amplified, digitized and classified as signals above and below average intensity values of the laser beam. The resulting stream of digitized random bits enters the Bit Pool. The same laser beam is split (for economical reasons) and is directed to a second detection unit detecting the *independent fluctuations* in intensity. They are amplified and digitized. This is similar to the PhRBG process. These digitized signals are also injected into the Bit Pool. The Bit Pool functions are discussed next.

## III. THE BIT POOL

THE Field Programmable Gated Array (FPGA) is a very convenient way to have a multitude of operations done in a dedicated programmable hardware instead of software, it brings an appreciable gain in speed: It is used in the Platform.

The Bit Pool in the KeyBITS Platform can be implemented in the same FPGA that performs the classification signals that output bit signals (digitized voltage signals) in the PhRBG. An adequate FPGA model with two inputs should be chosen: One for the bits coming from the PhRBG and the other one from the Noise Generator. The FPGA should have enough memory for processing the data necessary for the privacy amplification steps.

From the start, the Bit Pool contains a sequence of length $b_0$ of random bits secretly shared by A and B (RX receiving stations) (These bits were originally generated by the PhRBG as a secret coding sequence to create an initial $M$-ry set bases to code a fresh sequence of $a$ bits $(a_1, a_2, …)$ generated). This initial secret coding sequence $s$ is the information advantage legitimate users have over any attacker. *How this initial sequence is shared?* At some moment in time A and B have to have a first contact. Say that B is a member of a team directed by A and they meet for instructions. Or else, B is a client opening an account in bank A. Many other examples can be given.

Each $M$-ry basis gives a random voltage value that is added to the voltage representing one of the bits in $a$. To generate each basis $k \in M$, $m$ bits are needed: $M = 2^m$. Therefore, to send an *initial* sequence of $a$ fresh bits one needs $m \times a$ bits to form all bases to encode the bits from $a$. The *initial* sequence is then $b_0 = b_{0,1}, b_{0,2}, b_{0,3}, … b_{0,m \times a}$.

$b_0$ is partitioned in blocks of size $m$ and each block codes each bit of $a$. The first basis, for the first bit of $a$, is given by $(b_{0,1}, b_{0,2}, … b_{0,m})$ and so on. The specific $M$-ry basis *number $k$* corresponding to this set of random bits is

$$k_{0,0} = (b_{0,m}\, 2^{m-1} + \cdots + b_{0,2}\, 2^1 + b_{0,1}\, 2^0).$$

In the next paragraphs one general round of sharing $a$ bits will be described for which A and B already shared a sequence of bases bits $b$ of length $m \times a$ obtained from the former round.

By doing this, to any voltage representing a bit sent $a_i$ ($\in a$) another *voltage* $b_{iV}$ derived from the random bits $b_i$ is added:

$$a_i + b_{iV} \qquad (1)$$

The operation given by this sum (1) yields the voltage corresponding to the bit $a = 0$ or a=1 *plus* the basis voltage. The resulting signal is classical (can be assumed precisely known) and, as such, it can be identified precisely within the $M$-ry bases set: Although the specific basis *voltage* is random, it is generated by an algorithm connecting the random variable $k$ -one among $M$ possible values given by the shared set of $m$ bits. This algorithm could be the alternating function for $k$ values odd or even:

$$b_{iV}(k) = b_{max}\left[\frac{k}{M} - \frac{1-(-1)^k}{2}\right], \qquad (2)$$
$$(k = 0,1,2,3,… M-1),$$

where $b_{max}$ gives the voltage difference between bits spaced by even and odd values of $k$. The basis indicator $k$ will sometimes be dropped from the notation $b_{iV}(k)$ for simplicity.

Recall that $b_{iV}$ is a voltage value representing one basis in the $M$-ry coding and derived, through $k$, from the shared blocks of $m$ bits in $b$. In other words, $m$ random bits give $k$ and $k$ gives $b_{iV}(k)$.

Although $k$ is randomly chosen and $b_{iV}$ follows it, the mapping given by (2) is assumed to be known by the attacker – but not the specific $k$ value used. By studying this map and by guessing the possible bit value, the attacker could recover the basis used. Consequently, it is not enough for the users A and B to use this coding to protect a transmitted bit; the adversary could break it.

The aim of the noise generator shown in Figure 4 is to eliminate this recover operation: It creates random bits representing noise values $V_i$ that are added to $a_i + b_{iV}$ before sending the signal to B (or RX). This noise amplitude is judiciously chosen such that it covers several possible bases voltages. This *cloaking* effect frustrates the attacker on his aim for basis identification (or identification of the bit sent from A to B).

These cloaked signals are sent from A to B. The amplitude of this added noise $V_i$ is such that the noise covers several bases in the neighborhood of the actual basis used.

Bases values are separated by $2b_{max}/M$ and the noise amplitude is limited to be $\ll b_{max}$ to guarantee high fidelity in recovering bits by user B: If one recalls that users A and B know every basis sent, a simple subtraction of $b_i$ from $a_i + b_{iV} + V_i$, gives $(a_i + b_{iV} + V_i) - b_{iV} = a_i + V_i$.

The value $a_i + V_i$ gives the bit voltage sent $a_i$ plus a bit of noise $V_i$. As this noise is small ($\ll b_{max}$) the user B can easily recover the $a_i$ bit sent as well as the whole sequence $a = a_1, a_2, a_3, …$ .

By doing this the generated bit sequence $a$ sent to RX is now known by A and B. An example is given in the next section.

A and B now share a sequence of bits of size $a + m \times a$. The privacy amplification protocol (PA) in the Bit Pool [22] has to distill from this shared $a + m \times a$ bits two new sequences:
1) A random set $z$ of bits over which the attacker have **no** information whatsoever. This set $z$ will be used as keys for encryption, and
2) Another fresh sequence of size $m \times a$ to be used as the bases for the next round.

These steps will be discussed ahead.

## IV. CODING, DECODING AND NOISE

THE following simulations are used to illustrate the $M$-ry coding, decoding and the noise effect on a sequence of bits.

### A. M-ry coding

Figure 5 shows voltage value of the first 40 bases within a total of $M = 256$ bases. Typical voltage values and physical parameters are used. From now on the index $V$ will be dropped from $b_{iV}$ for simplicity – $b_{iV}$ is given by Equation (2). $b_{iV}$ is a voltage while the sequence of bits given the specific $M$-ry basis number $k$ ($\in M$) is derived from bits $b$.

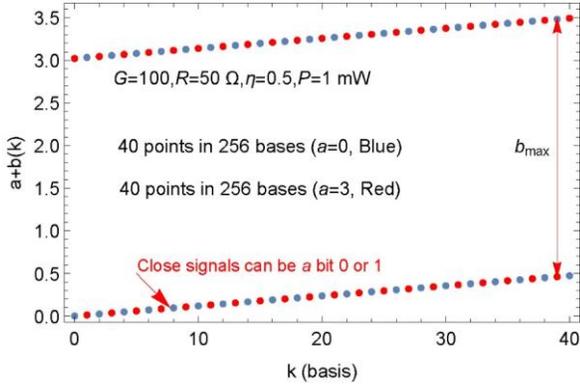

Fig. 5. An example of generation of bases levels with the M-ry coding. Values were set for bits 0: a=0V, and for bit 1: a=3V. Here $b_{max}$ was set arbitrarily to $b_{max}$=3V.

Although bases in Figure 5 are ordered in $k$ for illustration purposes, each $k$ is *randomly* chosen when coding a bit. The coding proceeds bit by bit. The number of levels $M$ used depends on the digitization supplied by the ADC. Typical values are $2^8, 2^{10}, \ldots$ . ADC financial cost increases with the number of bits $m = \log_2 M = 8, 10, \ldots$ and the speed offered. The FPGA applies the cyclic condition given by:

If $[a + b(k) > 2b_{max}]$ write $a + b(k) - 2b_{max}$, otherwise write $a + b(k)$.

This cyclic condition reduces the span of voltage values to a maximum value of $2b_{max}$. The minimum voltage separation between bases is $(2b_{max})/2$. See Figure 5.

### B. A fresh bit sequence

Consider a sequence of 64 bits:

$$a = 1001110000100111111011110010100$$
$$11010110011011001010011011111001010.$$

The circuitry map these bits as voltages values. Consider that $a_i = 0$ (bit 0) will be represented by $0V$ and $a_i = 1$ (bit 1) by $1V$. Figure 6 shows this sequence of bits.

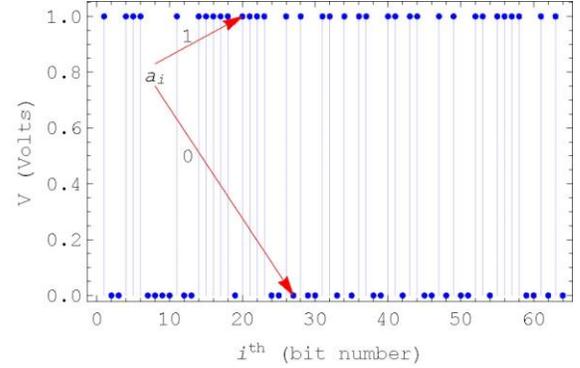

Fig. 6. A sequence of 64 bits represented as voltage values, where a bit 0 is arbitrarily represented by 0V and a bit 1 by 1V.

### C. M-ry coding a bit sequence

The $M$-ry coding uses the sequence of $m \times a$ bits secretively shared by A and B. For each bit $a_i$ a fresh sequence of $m$ bits define the basis $b_i$ for $a_i$. Assume, as an example, that this sequence of bases *numbers* chosen at random (levels within the $M$ bases) is given by $k$:

$$k = (96,115,151,82,129,242,96,79,58,195,224,8,208, \quad (3)$$
$$251,230,77,156,146,15,32,8,7,215,212,38,225,249,$$
$$106,84,9,254,252,202,219,223,86,84,173,238,237,247,$$
$$157,124,250,159,40,144,100,132,137,16,230,3,231,102,$$
$$132,112,51,193,54,253,62,102,246,128,64,72,136,43,$$
$$190,3,166,5,46,148,208,76,149,32,11,175,211,198,175,$$
$$248,86,26,99,61,168,34,105,47,137,121,10,64,126,52,$$
$$62,211,252,228,87,223,22,134,83,197,78,155,22,77,$$
$$150,110,167,199,28,236,182,94,240,206,9,96,155,$$
$$95,136,241,198,49,177,157,85,137,13,167,123,14,95,$$
$$198,85,0,84,196,53,145,51,22,32,194,196,84,2,42,$$
$$234,86,50,230,15).$$

The voltages corresponding to these bases $k$, call them $b$, are shown in Figure 7.

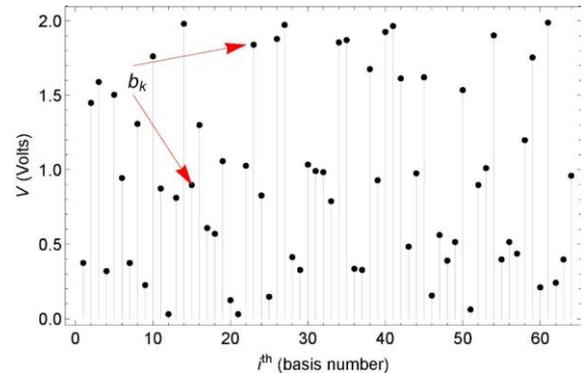

Fig. 7. A sequence of voltage values corresponding to random bases shown in list b are shown.

### D. Bits in coded bases

Figure 8 shows bits coded in the $M$-ry bases $a_i + b_i$ where the set $b_i$ is given by $k$ (3).

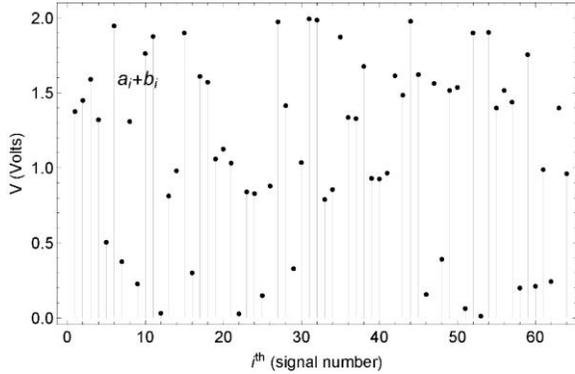

Fig. 8. A sequence of voltage values corresponding to a distinct basis coding every bit.

As the bases sequences are known only by A and B but not by the attacker, it may give the impression that only this secretly shared information could be enough to protect the in-transit information. However, the transmitted signals are classical and, in principle, any level of resolution of voltage signals could be assumed. An attacker examining signals coded with only this protection could use the generic algorithm for bases generation given by (2) and by comparison with the transmitted signals could infer both the bases and the bits sent.

To avoid that and safeguard the information, an extra layer of protection given by the optical noise is added to the coded signals as described below.

### E. Bits in coded bases with added noise

The red points in Figure 9 show the effect when noise is added to the coded signals (black points). Although the differences between noiseless and noisy signals are not large, the separation between black and red points is usually greater than the separation between two nearest bases $2b_{max}/M$. In this example, $M = 2^8$ was used, but if $M = 2^{10}$ was chosen, the separation between nearest bases decreases by 4 times. The desired net effect of the added noise is that the attacker cannot identify which basis or bit was used in any signal emission. The noise spans a voltage range around a coded bit sent, and this cloaks basis and bit sent.

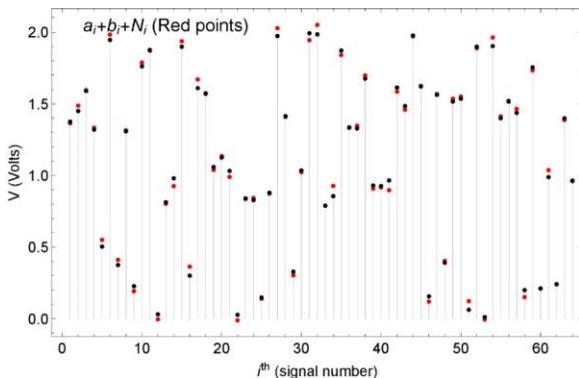

Fig. 9. Coded bit voltage (black points) added to noise voltage (red points).

### F. Subtracting the bases information shared by A and B

Receiver B now possesses the total signal sent by A shown in Figure 9. The attacker also has a recording of the same total signal but has no knowledge of the sequence of bases used. At the same time, both A and B know the bases used.

This is the *information advantage* that the users have over the adversary: B just subtracts the information on the $m \times a$ bases used (64 in this illustration) initially shared and obtains the result shown in Figure 10.

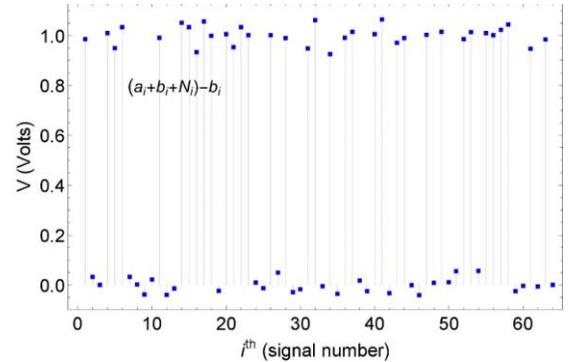

Fig. 10. B subtracts the information of the bases used and recovers the sequence of bits sent. Observe that these bits still keep the added noise signals but that does not prevents B to obtain the bit sequence. A simple rounding operation is needed (See Figure 6) for a perfect recovery of the bits $a$.

This result shows the importance of the *information advantage* created by A and B over the attacker thanks to the added noise signals. These noise signals present no formation rule and cannot eliminated or separated from the total transmitted signal.

## V. PRIVACY AMPLIFICATION PROTOCOL IN THE BIT POOL

A and B (or TX and RX) now possess the common knowledge of the starting shared key of length $m \times a$ used to create a first set of coding bases for $a$. A and B also shared a fresh key sequence of bits $a = a_1, a_2, \ldots$ (See Figure 6).

With these shared bits, the privacy amplification protocol (PA) creates a new set of bases (length $m \times a$) *and* a sequence of bits $z$ (smaller than $a$) for which the attacker has no knowledge and which could be used for one-time-pad encryption. The new set of bases provides a renewed secret shared by A and B (a fresh *information advantage*) to yield a new round of fresh keys sent from A to B. The key distribution process could then proceeds.

At this stage there are a few problems to be solved. The first one is that although the attacker have been frustrated at his intentions to obtain the bit sequence $a$ there is a probability, in principle, that he has obtained some correct bits in his trials. *Could this knowledge jeopardize someway the distribution process by allowing an increasing knowledge of the attacker about the sent keys?*

Advancing the answer without explanations it will be shown that the potentially acquired information by the attacker is completely negligible due to the recorded noise that cloak the signals. Furthermore, procedures in the PA protocol eliminates any residual information eventually left. Details are given in the Appendix XI.C.

As the intensity noise has a Gaussian probability around the voltage value assigned to $a_i + b_i$, a probability for the attacker to hit the right ``bit+$M$-ry basis'' exists. This probability for any single emission $i$ can be written $t_1$ and the probability for all sequence $a$ is $t = t_1 \times a$. If $t$ is less than one-half (in fact it results in a much smaller number) the PA protocol [5] teach us that the attacker's knowledge can be reduced to an infinitesimal amount.

In each round there are $a$ (bits) $+ m \times a$ (bases) in the bit pool, known by A and B, and over which the attacker may know a very small number of them. There are ways to eliminate this small statistical knowledge obtained by the attacker. A couple of possibilities are mentioned ahead. The choice of the protocol has to take into account the overhead imposed by the protocol itself because these operations reduce the overall throughput rate of the key distribution process.

## VI. UNIVERSAL HASH

DURING the data transmission ($a_i$ coded in $b_i$ bases plus noise) from A to B an instance of a universal hash function $f$ is set (it may include AES, Toeplitz matrices etc). A and B both apply the same hash operation to the sequence $a$ (bits) $+ m \times a$ (bases) to generate an output that keeps the same length as the input: This hash operation produces an extra randomization over the sequences possessed by A and B.

This hashed sequence passes by other transformation to eliminate the attacker's knowledge: The calculated number of bits $t$ potentially obtained by the adversary is used to *reduce* the initial number of bits $n = a + b$ to a smaller number $n - t$ (see [27]):

$$\{0,1\}^n \rightarrow \{0,1\}^{n-t}. \quad (4)$$

In other words, a number $t$ of bits is destroyed from the sequence $a + b$. An extra number of bits $\lambda$ is reduced as a security parameter [5]:

$$\{0,1\}^{n-t} \rightarrow \{0,1\}^{n-t-\lambda} \quad (5)$$

where $\{0,1\}^{n-t-\lambda}$ is the final sequence of bits, with length $r = n - (t + \lambda)$.

This reduced number $r$ of bits is grouped as follows:

$$\begin{aligned} r = n - t - \lambda &= (a + b) - t - \lambda \quad (6) \\ &= (a - t - \lambda) + b \\ &= (a - t - \lambda) + m \times a \\ &\equiv z + m \times a. \end{aligned}$$

The sequence of size $z \equiv a - t - \lambda$ is the sequence of *fresh bits* to be used for *encryption*. The sequence of size $m \times a$ will form the new bases $b_i$ for the next round of bit distribution.

**Perfect secrecy**: *Fresh keys $z$ were then acquired by and A and B without using a courier. The attacker has **no** information on $z$. This makes possible utilization of OTP encryption with secure keys $z$ and therefore achieving the **perfect secrecy level**.*

It should also be observed that even if an attacker could obtain a sequence $z$ for one round, say, from a known-plaintext attack or any other means, no past or future bit sequence is compromised; the PA protocol protects *each round independently*. Distilled bits in one round are uncorrelated with distilled bits from any other round. This is made possible by the process of *continuous injection of entropy* into the Bit Pool at every round – with fresh bits generated by the PhRBG.

Bennett [27] says that after reducing the initial number of bits from $n = a + m \times a$, ($m \times a$ initially shared to create bases and $a$ fresh bits) to $r = n - (t + \lambda)$, the amount of information that may be known by the attacker is given by the *Mutual Information $I_\lambda$*.

Corollary 5 (pg. 1920) in [27], gives the information in bits leaked to the attacker:

$$I_\lambda = \frac{1}{2^\lambda \times ln2} = \frac{1}{2^{n-(t+r)} \times ln2} \quad (7)$$

Observe that the high level of security achieved is due to the small amount of information leaked to the attacker – and this small amount was enforced by the physical noise added to the signals and the privacy amplification procedure.

Appendix XI.J gives the details and numerical estimates of the degree of security achieved as measured by the Mutual Information $I_\lambda$.

In conclusion, users A and B have a means to keep sharing in a secure way fresh keys from a continuously generating source (PhRBG). At each round a new set of bases of length $m \times a$ is generated and, a set of $z$ are available for encryption. The size $z$ obtained is reduced from the original size $a$ by $(t + \lambda)$. This reduction is the *overhead* of the process. Numerical examples will be given in Appendix XI.I.

Figure 11 summarizes the privacy amplification protocol steps.

| $m = log_2 M$ | [$m$ = number of bits to create one basis among $M$] |
| $m \times a$ | [preshared "key" between A and B] |
| | [$a$ is the number of fresh bits to be shared by A and B] |
| | For each bit sent a basis of $m$ bits is used. |
| $n = a + m \times a$ | [$n$ is the total number of bits possessed by A to B] |

A shuffling procedure is applied over the $n$ bits with B. An identical operation is done on A.

$t$ = bits probabilistic obtained by attacker after bits $a$ are sent [$t = t_1 \times a$, and $t_1$ is the success probability for an attacker to get one bit after this bit is sent].

A and B discard $t$ bits: $n \rightarrow (n-t)$. Bits left: $\{0,1\}^n \rightarrow \{0,1\}^{n-t}$

$\lambda$ = safety parameter (extra bits destroyed) $\rightarrow r = n - (t + \lambda)$   [$r$ is the final number of bits kept]

Arrange $r$ as $r = n - t - \lambda = (a - t - \lambda) + m \times a \equiv z + m \times a$.

$z$ is the final random sequence of bits for encryption and $m \times a$ form a new set of bases  ($z < a$)

Mutual Information between the attacker and B: $I_\lambda \equiv I(n,t,r) = \frac{1}{Log_e(2)\, 2^{n-t-r}}$

Fig. 11. Summary of the privacy amplification steps.

## VII. ONE-TIME-PAD ENCRYPTION

ONE-TIME-PAD encryption can proceed over any channel with perfect secrecy and with the speed given by the generation rate of $z$. This will be roughly the bit generation rate of the PhRBG minus the overhead for PA occurring in the Bit Pool.

*A. Decentralized encryption*

Figure 12 shows a TX station and up to N RX stations. The one-time-pad encryption works between one-to-one users (say TX and one RX) but can also work between one-to-N as (say TX to N RX) or as well as any arbitrary number of users in the network.

A decentralized possibility for encryption based on a same set of encryption of keys may be particularly useful for a team with N members [28].

This decentralized use proceeds as follows. Assume that TX has distributed a certain number of coded information with added noise random keys to N users (that form a team) and that software will automatically apply the same distillation process given by the PA protocol (It is assumed that these users pre-shared a sequence of random keys to form a set of $M$-ry bases).

All team members will obtain the same set of fresh encryption keys $K$ ($K$ may be a large number of sequences $z$). See Figure 12.

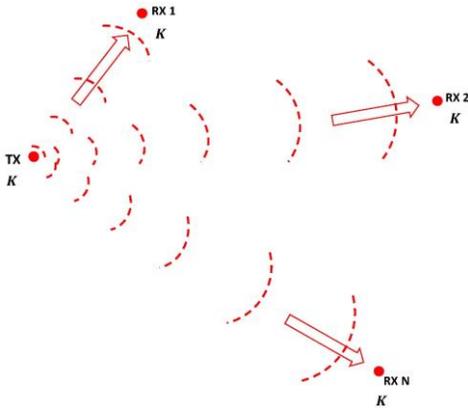

Fig. 12. TX sends to N RX a sequence of coded bits with added noise. All stations perform the PA operations and end up with a sequence of fresh bits K.

Assume these $K$ bits are arranged in a square matrix form. See left side of Figure 13 as an example. Lines can be enumerate from top to bottom sequentially 1,2,3, ... $\sqrt{K}$.

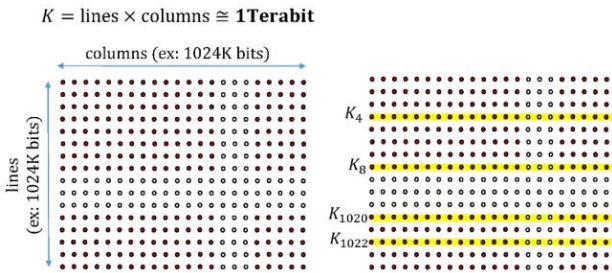

Fig. 13. *At left, arrangement of the K bits in a square matrix. At right, an example of randomly choosing lines in the matrix.*

Consider that one user "$n$" wants to send to user "$m$" a message with length equal to one line in that $K$ matrix. User "$n$" chooses randomly 20 lines in the $K$ matrix (see right side of Figure 13 and applies an XOR operation over these 20 lines:

$$K_4 \oplus +K_8 \oplus ... K_{1022}, \quad (5)$$
$$(20 \text{ sequences})$$

This obtained XOR sequence is the sequence of bits to encrypt bit-by-bit the message $M$:

$$C = (K_4 \oplus +K_8 \oplus ... K_{1022}) \oplus M \quad (6)$$

The encrypted message $C$ is sent to user "$m$" with a header containing the numbering of the encryption lines (that were "XOR"-red). Decryption is easily done by "$m$" because he holds the original sequence $K$. For a different number of lines an obvious extension of the procedure can be employed.

The attacker does not know $K$ nor even the content of any line. At most the attacker could have obtained the order of lines in the header of the encrypted message $C$.

The collision probability to have the same line chosen in another encryption by any user can be calculated as well as to have a collision of all the lines. These (very low) probabilities are calculated in Appendix XI.K .

In the decentralized use of a batch of keys, one estimate is that after multiple uses all keys in the total number of keys would have been used at least once, the process can start from the beginning. In case of one-to-one encryption the used keys could be discarded immediately after use while for the decentralized case keys they can be discarded only after a renewing process for all users happens. Figure 14 illustrates a possible one-to-one key distribution and encryption and a one-to-N key distribution with decentralized encryption.

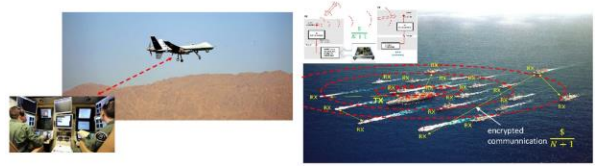

Fig. 14. Left side - A ground station distribute keys to a drone that by its turn sends encrypted images continuously to the ground station. Right side - A central unit TX send keys to N users. Decentralized encryption assures secure communication among the team. The technology cost is divided among users.

Many applications can be derived from the keyBITS technology. Figure 15 shows an example of protection of transportation structure that could be compromised, adversely affecting air traffic if maintenance data falls into the hands of terrorists.

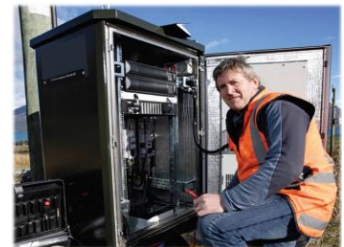

Fig. 15. Securing maintenance data transmission for transportation structures in the nation.

*B. OTP Graphical Interface*

A very easy-to-use Graphical Interface was developed so that the user only has to point to file to encrypt and to path to the stored key. All random choices are used and encryption is done in a fast way and made disposable to the user. It performs the tasks of Encryption and Decryption using the decentralized protocol. All operations are rapidly accomplished in the background. Figure 16 shows the basic interface.

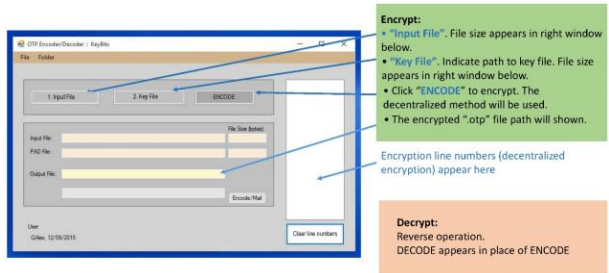

Fig. 16. Friendly-to-use Encode/Decode Graphical Interface.

## VIII. COST COMPARISONS BETWEEN KEYBITS AND QKD

ALTHOUGH costs of installed QKD systems are not readily-available, rough estimates can be made based on specific cases. The Quantum Key Distribution case known as Tokyo QKD Network [29] is shown in Figure 17.

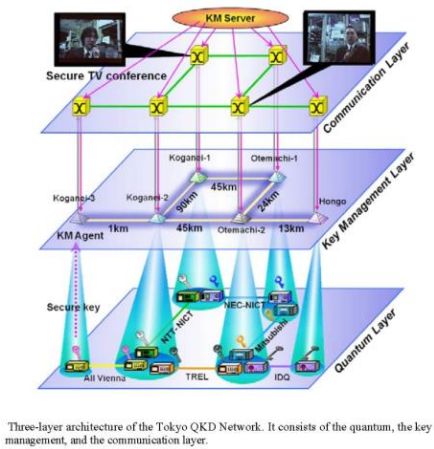

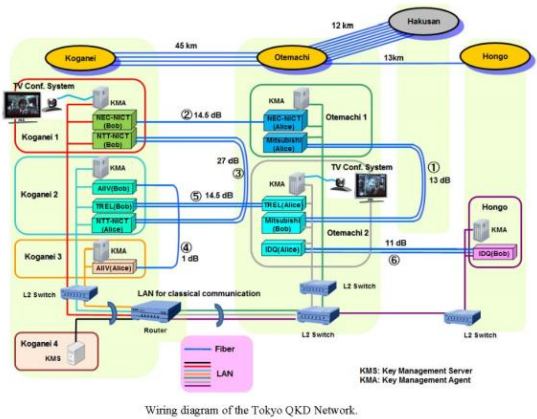

Fig. 17. Tokyo QKD network.

This network has 6 stations. The stations use close to 289 km of optical fibers, 6 pairs of QKD stations, at least 3 switches, 1 router and transmission cables. Infrastructure costs in U.S. dollars can be estimated: $US\$ \ 62 \times 10^3$/km for the cost of a km of fiber, $US\$ \ 2 \times 10^5$ for each pair of QKD Alice/Bob stations, and about $US\$ \ 50 \times 10^3$ for computers, routers, switches and cables. A rough equipment cost of $US\$ \ 3.2 \times 10^6$/station is obtained: Total of $\$ \ 19 \times 10^6$. The cost is linear with the number of stations starting with a minimum of two stations. Estimates for operational expenses is dependent on organization-specific factors and is not included here. Accounting for these can be done with a cost-benefits analysis.

The KeyBITS platform can operate station to station with independent connections or in the decentralized mode with one Platform connected to $N$ receiving stations. When KeyBITS is considered as an independent entity, the cost is linear for a group of two end-points. In the decentralized case the main cost of the Platform ($C_P$) is roughly calculated as $C_P/(N+1)$ Figure 18 depicts the cost decay as a function of the number of users N. It is assumed that the users already have PCs or other communication devices with installed KeyBITS privacy amplification programs.

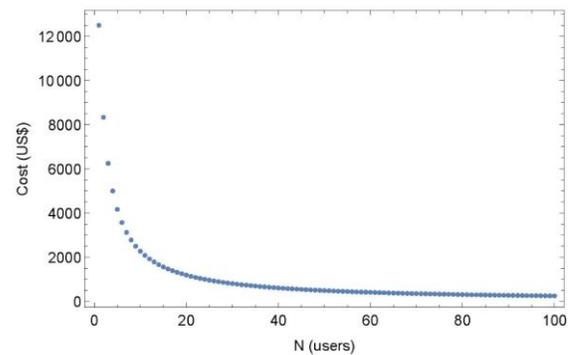

Fig. 18. Costs as a function of the number of decentralized users N. Platform cost estimated at $US\$25 \times 10^3$.

With a rough estimate of $US\$ \ 25,000$ for each Platform, two stations with one Platform each, costs $US\$ \ 50,000$ (to be compared with the QKD cost of $US\$ \ 3.2 \times 10^6$). For the decentralized configuration consisting of one Platform and 100 users, the average per user cost would be $US\$ \ 248$.

These KeyBITS costs are based on the use of commercial components procured at unitary prices. Components procured in bulk will realize appreciable price reduction. Furthermore, miniaturization of the Platform for a large chip size could produce an even higher cost reduction.

## IX. CONCLUSIONS

THE KeyBITS Platform for Key Generation, Secure Distribution and Encryption was based on several novel ideas. Proven security properties of KeyBITS that outperform other encryption methods are:

- Physically generate keys, using the quantum fluctuations in a laser beam. No generation algorithm exists.
- Proven secure distribution of keys. Signals are protected by physical noise and privacy amplification procedures. Each sequence of distilled keys $z$ has no correlation with past or future sequences of distilled keys; the distillation is compartmentalized round by round.

- Secure transmission of data, images, voice using one-time pad encryption. Distinct services are possible over the same Platform: from military uses to IoT, protection of financial or medical data transmission, and so on.
- Continuous and fast key distribution without use of couriers.
- The optoelectonics system is stable with no use of interferometry.
- Key generation is fast and dependent only on the speed of the electronics. The physical principles used accept higher speeds.
- Miniaturization is possible for a large chip size.
- Can be built with commercial parts.
- Easy-to-use Graphical User Interface to encrypt/decrypt.

The KeyBITS technology does not use Quantum Key Distribution (QKD) protocols: It was developed to be faster and cheaper than QKD, without restrictions for long range communications. Transmitted signals are classical, telecom standard, but carry recorded optical noise of quantum origin to create a physical cloak that hides the signal bits transmitted.

The KeyBITS Platform starts generating cryptographic keys in a fast ($> 2G\text{bit}/s$) process, continuously, by using a novel, patented technology that resulted in the KeyBITS Physical Random Bit Generator (PhRBG). The entropy source of bits are the quantum fluctuations (optical shot noise) of a laser field. The bit generation rate is above 2Gbit/sec with the current electronics and only bound by electronic circuitry. Any advances in electronics can be incorporated in the system because the quantum fluctuation process is very broadband (white noise). The overhead for the distillation of encryption bits is due to the $M$-ry coding used and the privacy amplification protocol; the corresponding speed overhead cannot be eliminated but it can be minimized by using a faster electronics.

Another unique, novel KeyBITS feature is the decentralized encryption for multi-users. After first sharing identical keys for N users, the technology allows ongoing decentralized encrypted communication among the N users.

This novel generator (PhRBG) can function as stand-alone equipment and as such can be marketed independently of the Platform.

It was demonstrated that the one-time-pad encryption can be revived with the KeyBITS technology. A system to give perfect secrecy for all in-transit communication, fast and low cost compared with QKD technologies is now within our reach.

## X. Appendix

### A. Perfect secrecy - One Time Pad

Shannon [30] discussed the entropy $H$ of an information source of variables $x$ ($x \in X$) characterized by a set of probabilities $P(x)$

$$H(x) = -\sum_{x \in X} P(x) \log_2 P(x) \qquad (7)$$

The symbol $X$ can be attached to a sequence of random variables $K$, to a message $M$, an encrypted text $C$ and so on. For binary variables, $H(x)$ can be measured by "number of bits". The conditional entropy $H(X|Y)$ gives the uncertainty in $X$ when $Y$ is known:

$$H(X|Y) = -\sum_{x \in X}\sum_{y \in Y} P(x,y) \log_2 P(x,y), \qquad (8)$$

where $P(x,y)$ is defined as

$$P(X = x|Y = y) = \frac{P(X = x, Y = y)}{P(Y = y)}. \qquad (9)$$

An encryption system with key $K$, to offer some uncertainty to an attacker, has to conform to

$$H(M|C) \leq H(K|C). \qquad (10)$$

An encryption system offering perfect secrecy must meet this parameter:

$$H(M) \leq H(K), \qquad (11)$$

It must have a state such that the randomness of the key is not less than the randomness of the message.

The Mutual Information $I(X;Y)$ between random variables $X$ and $Y$ measures the information (in bits) obtainable for the variable $X$ after $Y$ is known. It is defined by

$$I(X;Y) = \sum_{x \in X}\sum_{y \in Y} P(x,y) \log_2 \frac{P(x,y)}{P(x)P(y)}. \qquad (12)$$

Vernam's cipher (in binary) where the cipher text $C$ is achieved by the addition (modulus 2) of the message $M$ ($|M| = N$) and the key $K$ ($|K| = N$): $C = M \oplus K$, gives the Mutual Information (please note that $K$ is assumed uniformly distributed):

$$I(M;C) = \sum_{m \in M}\sum_{c \in C} P(m,c) \log_2 \frac{P(c|m)}{P(c)} \qquad (13)$$
$$= \sum_{m \in M}\sum_{c \in C} P(m,c) \log_2 \frac{1/2^N}{1/2^N} = 0.$$

The above gives the assurance that if the attacker comes into possession of the transmitted encrypted text $C$, nothing with certainty about the message $M$ is obtained.

The quantities $H(X)$ or $I(X;Y)$ are expectation values based on probabilistic calculations. In other words, they may indicate the probabilistic outcome or number of bits possibly obtainable. In this sense the result $I(M;C) = 0$ will be understood as a negligible number of bits obtained using accepted probabilistic calculations.

## B. Physical Random Bit Generator (PhRBG) - details

The PhRBG is an opto-electronic device designed to continuously generate and supply bits to when high speeds are required. The physical principle involved, quantum vacuum fluctuations that produce the optical shot-noise, is not bandwidth limited; device speed can be adapted to all electronic improvements. Important differences between the PHRBG and other quantum random bit generators include no need for interferometry and that a single detector is used whereas some other generators require two. This architecture yields a time-stable system.

The PhRBG is currently implemented with commercial-off-the-shelf components including low cost amplifiers (See G in Figure 1). These amplifiers have a frequency dependent gain profile (a monotonous high gain at low frequencies) that introduces a low frequency bias in the bit generation.

### a) Laser Power Spectrum

Figure 19 shows typical spectra from a diode laser of the type utilized in the PhRBG.

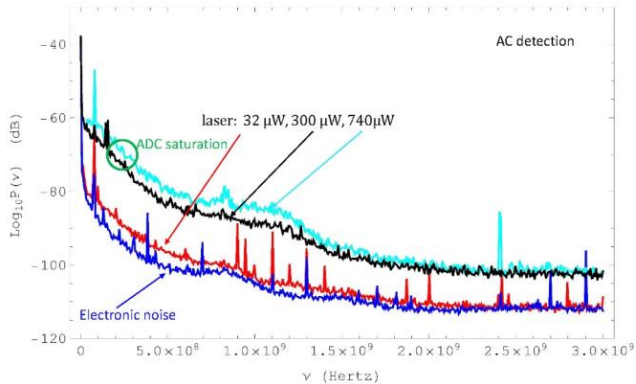

Fig. 19. Laser power spectrum with different light intensity levels. The spectra are not flat as ideally desired. Several filtering and external features can be seen. Filtering occurs at several stages: the optical filtering at the detector's glass window (passage band from infrared to UV) and at distinct frequency dependent impedances in all electronic stages including at the amplifiers. The filtering produces a final frequency response increase at low frequencies. These distortions, or deviations, from an ideal flat frequency response enhance the occurrence of slow phenomena as compared to the fast ones.

Correction of these frequency distortions can be done with a more-elaborate electronic circuitry (which is costly) or accomplished by an auxiliary randomization using, for example, a Linear Feedback shift Register (LFSR) -an inexpensive solution to break the occurrence of systematic features.

To compensate for this bias without increasing costs the LFSR is used in series with the bit output to produce an extra randomization. This breaks the long sequences of repeated bits, which are expectedly more rare. The process does not reduce the speed of the PhRBG.

As currently implemented, the PhRBG operates at ∼ 2.0 Gbit/sec and passes all randomness tests to which it was submitted, including the NIST suite described in [23].

Saturation signals are another feature related to the resolution presented by the ADC. Whenever the voltage related to the intensity fluctuations rises above the maximum allowed ADC voltage, a null or saturation response occurs. From the other side, if the noise is below the voltage corresponding to the resolution of the ADC the system, no faithful response is obtained as well.

Recall that the Poissonian photon fluctuation with an average intensity $\langle n \rangle$ presents a photon number fluctuation $\sigma_n = \sqrt{\langle n \rangle}$. The relationship of noise-over-signal becomes

$$\frac{\sigma_n}{\langle n \rangle} = \frac{1}{\sqrt{\langle n \rangle}} \ . \tag{14}$$

That is, for high $\langle n \rangle$, the ratio $\sigma_n/\langle n \rangle$ can become exceedingly small, below the sensitivity of the ADC.

Correction of the problem can be done with an ADC of increasing resolution: the spectra shown in Figure 19 were achieved with an ADC of $2^8$ levels. An ADC with $2^{10}$ levels would present less saturation and so on. Again, correction is a question of the cost/benefit ratio.

### b) Fibonacci LFSR

Figure 20 shows a Fibonacci LFSR used to produce the auxiliary randomization. It is sequenced to operate immediately after the electronic output of bits (IN gate) and output the mixed signals (OUT gate) and without speed loss.

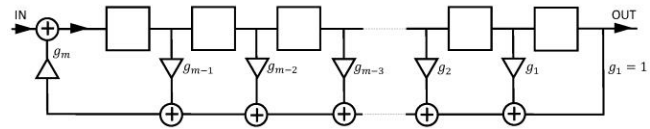

Fig. 20. The raw random bit stream employs a Fibonacci's LFSR as an auxiliary step to break bias created by the amplifier gain profile of the. There is no speed decrease.

The resulting bit output passes all NIST and other randomness tests. Figure 21 exemplifies main features of the bit output.

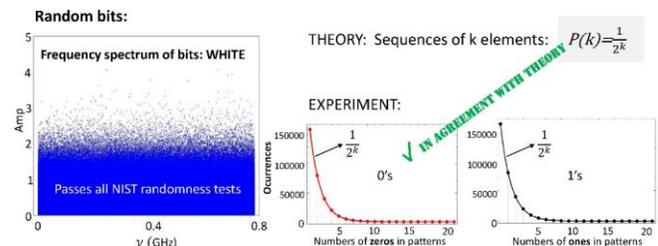

Fig. 21. Left - Frequency spectrum of the randomized bit output and examples of repeated sequence analyses. Transforming (0,1) sequences onto (-1,1) sequences allows the spectrum analysis with "white-noise" character of the output signals. Right - Sequences of repeated elements obey the expected probability $P(k) = 1/2^k$: Solid line is theory and points the experimental data.

For a distribution where the probabilities to occur a 0 or 1 are equal, $p = 1/2$. It is expected that the probability of a sequence of $k$ identical bits (either 0 or 1) occurring is $P(k) = 1/2^k$. If one changes basis 2 to basis "$e$" the equation is

$$P(k) = \frac{1}{2^k} = e^{-k \ln 2} \cong e^{-0.693\, k} \ . \tag{15}$$

Data in the right side of Figure 21 were fitted to $p(n) = c \times e^{\ln 2^{1-\varepsilon}\, n}$, where $\varepsilon$ will indicate a depart from the distribution $P(k) = 1/2^k$. Dots are obtained from $\sim 1.3 \times 10^6$ bits and the solid lines shows the fit to $c = 319880 \pm 193$ and $\varepsilon = -0.003 \pm 0.002$.

In summary, some corrections to deviations from ideal conditions, such as a non-flat gain G, could present different costs. Usually, a more inexpensive solution is chosen to achieve the desired goals.

The raw data for the histograms are given by lists $L_1$ and $L_2$:

$L_1 = \{\{1, 159676\}, \{2, 79651\}, \{3, 40253\}, \{4, 20017\}, \{5, 9864\}, \{6,4960\},$
$\{7, 2567\}, \{8, 1239\}, \{9, 623\}, \{10, 313\}, \{11, 156\}, \{12,59\}, \{13, 37\}, \{14, 21\},$
$\{15, 9\}, \{16, 8\}, \{17, 3\}, \{18, 4\}, \{19, 1\}, \{20, 0\}, \{21, 0\}\}$

$L_2 = \{\{1, 159805\}, \{2, 79964\}, \{3, 39766\}, \{4, 20021\}, \{5, 9892\}, \{6, 4962\},$
$\{7, 2488\}, \{8, 1306\}, \{9, 630\}, \{10, 336\}, \{11, 148\}, \{12,71\}, \{13, 42\},$
$\{14, 10\}, \{15, 11\}, \{16, 6\}, \{17, 2\}, \{18, 0\}, \{19,1\}, \{20, 1\}, \{21, 1\}\}$

Observe that the deviation parameter $\varepsilon$ gives an estimate of the deviation from the ideal expected behavior. In this fit it is shown to be very small, an estimate of the randomness associated with the generated bits.

### C. Privacy amplification and perfect secrecy

The perfect secrecy goal is achieved by associating random optical signals in the coded transmitted signals and a somewhat conventional PA protocol that relies heavily on the effects of the random noise signals. The two parts will be discussed separately below.

The random noise will delimit the amount of information that an attacker could obtain from the $M$-ry coded bit superposed with noise. This leaked information will be represented by $t_1$.

The PA software eliminates these residual statistical data that could have been acquired by the attacker. Sequences of fresh random bits $z$ for encryption results; the attacker can obtain no practical information.

### D. Detected Physical Signals - *from photon numbers to voltages*

Photon (or light) detection in general has a long history (See, for example [31]). Some basic formulae that aid in describing the physical signals are presented here. These represent each bit inside the KeyBITS Platform that will be transmitted between encoder and decoder.

While all collected formulae won't be demonstrated here, these can be considered intuitively acceptable. For those wanting to consult a single reference to check derivations for these formulae, please see Sections F through H in [32].

The circuitry shown in Figure 1 can be represented by the diagram in Figure 22.

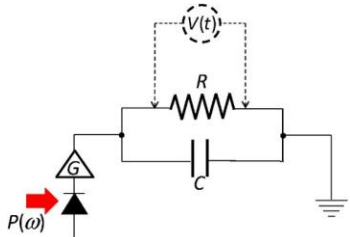

Fig. 22. Equivalent circuitry representing the detection components of the PhRBG. R and C are the equivalent impedance elements (resistance and capacitance) of the voltage detection equipment represented by the ADC. The voltage V represent the analog voltage signals that will be discretized by the ADC in $M$ voltage levels.

The light beam interacting with the detector, with power $P$, contains an average photon number $\langle n \rangle$ per unit of time (perhaps, for example 1sec): $\langle n \rangle_1$. This gives

$$P = \langle n \rangle_1 \hbar \omega_0, \quad (16)$$

where $\omega_0$ is the laser frequency in radians/sec.

The photon number statistics for a coherent laser beam is Poissonian:

$$p(n) = \frac{e^{\langle n \rangle} \langle n \rangle^n}{n!}. \quad (17)$$

The process inside the detector creates a photocurrent $I_e$ (here $e$ refers to the electric charge) that, by its turn, presents a Gaussian distribution:

$$P(I_e) = \frac{1}{\sigma_e \sqrt{2\pi}} e^{-\frac{(I_e - \langle I_e \rangle)^2}{2\sigma_e^2}}. \quad (18)$$

Crossing the $RC$ impedance the electrons generate a voltage $V$ that also follows a Gaussian distribution:

$$P(V) = \frac{1}{\sigma_V \sqrt{2\pi}} e^{-\frac{(V - \langle V \rangle)^2}{2\sigma_V^2}}. \quad (19)$$

The detection theory provides the value

$$\langle V \rangle = RGe\eta\langle n \rangle_1. \quad (20)$$

where $G$ is the amplifier gain, $e$ is the electric charge, and $\eta$ is the detector efficiency. It also gives

$$\sigma_V = \sqrt{\frac{R}{2C}\left(G^2 e^2 \eta \langle n \rangle_1 + \frac{2 k_B T_K}{R}\right)}, \quad (21)$$

where $k_B$ is Boltzmann constant and $T_K$ is the temperature in degrees Kelvin. The second term gives the voltage fluctuations due to thermal effects in the circuit.

From the $M$-ry coding of bits and the addition of noise introduced in Section IV, conditions for protection and recovery of in-transit signals by the legitimate users can be defined.

The basic assumption is that the signal bit as well as the bits added for the noise were created by uncontrollable optical fluctuations and not by the background noise caused by thermal effects. This can be assured by the choices of $\langle n \rangle_1$ and G. Noise caused by thermal effects can be technically reduced by using lower temperatures, while the inherent optical noise in unaffected.

The physical fundamental equations for many estimates of the security are given by Eqs. (19) to (21).

### a) Optical fluctuations stronger than thermal induced electrical noise

Eq. (21) shows that for optical fluctuations bigger than the fluctuations imposed by the thermal part one should have

$$G^2 e^2 \eta \langle n \rangle_1 \gg \frac{2k_B T_K}{R} \quad (22)$$

that can be satisfied by adjusting $G^2 \langle n \rangle_1$. It should also be recalled that the ADC resolution, or "Least Significant Bit" ($LSB$, measured in voltage), given by the separation of its discrete levels, should be smaller than the optical fluctuation voltage

$$Ge\sqrt{\frac{R}{2C}\eta\langle n\rangle_1} > LSB , \quad (23)$$

so that the voltage fluctuation is correctly detected.

To calculate the $LSB$ voltage one has to know the full-scale voltage ($FS$) range of the ADC used and its number of discretization levels $M$. Taking, for example, $FS = 5V - (-5V) = 10V$, $LSB = 10V/M$, where $M$ is created by a number of bits $m$: $M = 2^m$. $m = (8, 10, 12, 14, ...)$, one obtains

$$\frac{10V}{2^8} = 39mV , \quad \frac{10V}{2^{10}} = 9.8mV , ... . \quad (24)$$

### E. Physical conditions for secure transmission of signals

#### a) Added optical noise smaller than bit separation voltage

The digitized minimum voltage separation in the ADC is related to the specific number of levels $N$ of the ADC and is given by the ADC number of the bits (8 bits → $2^8$ levels, 10 bits → $2^{10}$ levels, etc) and to the maximum voltage span $V_{max}$ allowed. The minimum separation is then $\Delta V = V_{max}/N$. Eq. (2). It defines the voltage $b_{max}$ that separates bits 0 and 1. For example, $b_{max}$ could be set at $V_{max}/2$. This condition gives

$$2\sigma_V \ll \frac{V_{max}}{2} , \text{ or } 4\sigma_V \ll V_{max} \quad (25)$$

#### b) Added optical noise covers many bases

The total noise around one coded bit sent must be much bigger than the separation between nearest coded bits in the $M$-ry bases:

$$\frac{2\sigma_V}{\left(\frac{V_{max}}{M}\right)} \gg 1 , \text{ or } 2M\sigma_V \gg V_{max} . \quad (26)$$

#### c) Conditions obeyed

The set of practical basic conditions to be obeyed allowing security in the key distribution process are given by Eqs. (22), (23) and (24).

Figure 23 shows the surfaces $2M\sigma_V$, $\langle V \rangle$ and $4\sigma_V$ as a function of gain $G$ and optical power $P$.

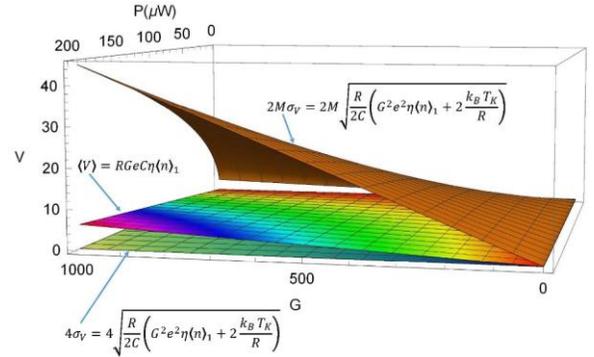

Fig. 23. Optical and electrical conditions obeyed by the keyBITS Platform. V (in Volts) is the electrical output due to bit, bases and noise signals. Here it was used $V_{max}=\langle V \rangle$. Observe that conditions $2M\sigma_V \gg \langle V \rangle \gg 4\sigma_V$ are obeyed, as required for secure transmission. Parameters were set at $M = 1024, \eta = 0.5, T_K = 300K, R = 50\Omega, C = 1 \times 10^{-12}F$. The electrical noise due to thermal effects gives $\sigma_{thermal} = 64\mu V$.

For an ADC with a maximum span of voltage of $V_{max} = 20V$, one could choose the voltage for a bit 1 equal to $V_{max}/2 = 10V$. This value can be set to equal the $\langle V \rangle$ output from the ADC for a given set of parameters $M, G, P$. This chosen value has also to obey the conditions (25) and (26).

Figure 24 exemplifies this choice in a log scale for clarity. The surfaces $\log_{10}\langle V \rangle$, $\log_{10} 2M\sigma_V$, $\log_{10} 4\sigma_V$ and $\log_{10}(V_{max}/2)$ are shown.

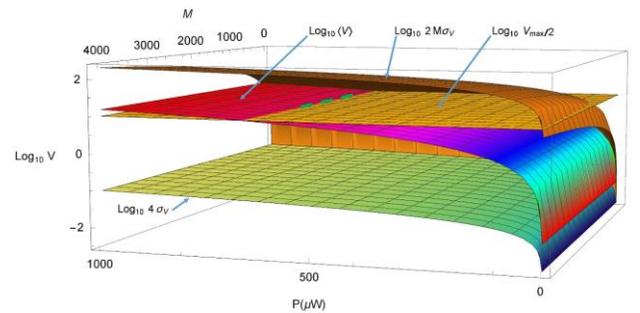

Fig. 24. Surfaces $log_{10}\langle V \rangle$, $log_{10}2M\sigma_V$, $log_{10}4\sigma_V$, and $log_{10}V_{max}/2$.

Value $V_{max}/2 = 10V$ is located at the crossing of the surfaces $\log_{10}\langle V \rangle$ and $\log_{10}V_{max}/2$, see dashed green line. In particular, for $P = 662\mu W$ and $M = 2^{10} = 1024$ the result is $\log_{10}\langle V \rangle = V_{max}/2 = 10V$ and $2M\sigma_V - V_{max}/2 = 31V$ and $\frac{V_{max}}{2} - 4\sigma_V = 9.9V$. Conditions (25) and (26) are obeyed with these settings.

### F. The attacker probability of error

It is opportune to comment that in the transmission of signals by an optical channel, such as an optic fiber with *phase* modulation, the variance of the noise component depends on $1/\langle n \rangle_1$ instead of $\langle n \rangle_1$ (See [5]). The *voltage* fluctuations used in the Platform are proportional to $\langle n \rangle_1$. Phase and amplitude fluctuations do *not commute* in the quantum domain (although they are not conjugate variables in the canonical sense). Therefore, their behavior variances are similar to complementary variables (or Fourier conjugate variables in the classical domain); this is the reason for a distinct dependence of the variances on $\langle n \rangle_1$.

The signals in the Platform are classical, and the fluctuation increasing with the power intensity seems to indicate that strong power would be ideal. But detection of the fluctuations are bound by the resolution of the ADC: For higher intensity ($\propto \langle n \rangle$) the ratio $\sigma_n / \langle n \rangle = 1/\langle n \rangle$ is too small; the ADC becomes insensitive -although $\sigma_n$ is larger. The ADC cost increases with a better resolution. The quantum phase modulation case [5] and the wireless case with the intensity variable lead to distinct calculations for the probability of the desired variables - one is necessarily quantum and the other one uses classical probability functions.

Fig. *9. Coded bit voltage (black points) added to noise voltage (red points).*
, previously shown, gives an example of the sequence of signals carrying voltages corresponding to bits+coding bases+noise voltages.

Figure 11, previously displayed, shows the decoding applied by the legitimate user by simply subtracting the known bases sequence. In that example $b_{max}$ was set to $1V$.

The attacker does not know this sequence, but consider this situation: Assume he receives the red dot signal $(a_i + b_i + V_i)$ seen in Figure 10. Take this voltage close to the correct noiseless signal as seen in Figure 6. The attacker knows that the noise would have randomly displaced the voltage value, but he also knows that the coding procedure creates alternate bits to the closest signals –see Figure 5. There are many signals with the same bit (see red and blue dots in Figure 5). By chance he may have hit the right bit. If the bit sent was a blue one, he could hit it by chance if the signal was coming from *any* to the blues ones or, he could err if the bit was a red one (any of the red points). The two sets are displaced from each other by one (small) separation $V_{max}/M$.

Designating $P_R$, the probability of hitting the *right* configuration of bits (blue or red set of points) that represent the bit sent and by $P_W$ the probability to hit the *wrong* configuration, it is known that one of the cases will occur. Therefore $P_R + P_W = 1$.

It is also known that the noise has a Gaussian distribution that will be centered on the signal $a_i + b_i$, and that the probability of success will be maximum at this point. $P_R$ and $P_W$ can be written in a normalized form as

$$P_R = \frac{\sum_{j=0 \, (j \, even)}^{M-1} e^{-\frac{(jV_{max}/2M)^2}{2\sigma_V^2}}}{\sum_{j=0}^{M-1} e^{-\frac{(jV_{max}/2M)^2}{2\sigma_V^2}}} \quad (27)$$

$$P_W = \frac{\sum_{j=0 \, (j \, odd)}^{M-1} e^{-\frac{(jV_{max}/2M)^2}{2\sigma_V^2}}}{\sum_{j=0}^{M-1} e^{-\frac{(jV_{max}/2M)^2}{2\sigma_V^2}}} \quad (28)$$

At this point one should realize that the maximum probability of *error* in a coin tossing is not zero but 1/2. As a counterpart, the minimum probability of *success* starts at 1/2 and not at 0. This minimum for the probability of success defines

$$P_s = \frac{1}{2} + \frac{1}{2}(P_R - P_W) \quad (29)$$

The first 1/2 accounts for the minimum value of $P_s$ and the second considers one group of points - upper or lower line in Fig. *5. An example of generation of bases levels with the M-ry coding. Values were set for bits 0: a=0V, and for bit 1: a=3V. Here $b_{max}$ was set arbitrarily to $b_{max}=3V$.*
. The minimum value of $P_s$ ($= 1/2$) would be the maximum for the probability of error $P_e$. This leads to

$$P_e = 1 - P_s \quad (30)$$

Figure 25 shows $P_s$ and $P_e$ for a range of parameters.

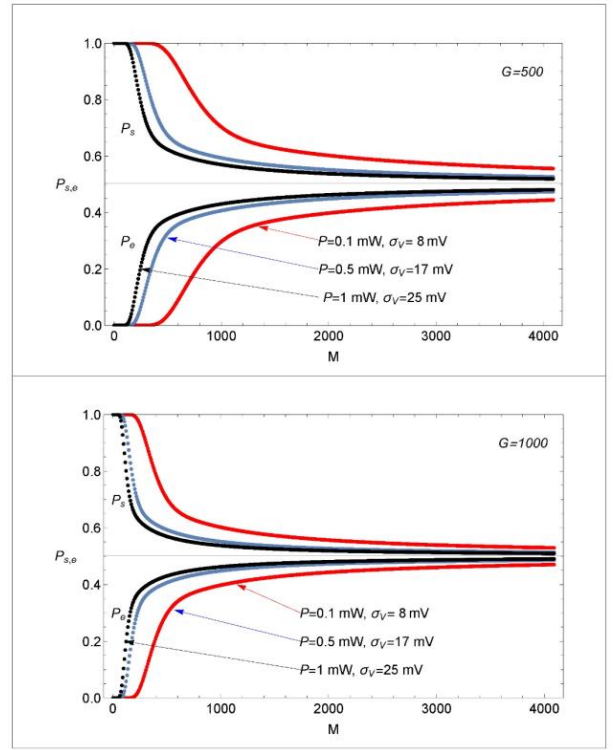

Fig. 25. Probabilities for error of an attacker, $P_e$ and success $P_s$, for a set of parameters and $G = 500$ and $G = 1000$. $\sigma_V$ values are obtained from Eq. (21) with parameters and constants $C = 1 \times 10^{-12}F$, $e = 1.6010 \times 10^{-19} Coulomb$, $k_B = 1.38 \times 10^{-23} J/K$, $R = 50\Omega, \eta = 0.5, T_K = 300K$. The maximum error that the attacker can make is to achieve the level of 1/2 equivalent to a pure coin tossing for the guessing of each bit. This 1/2 coin tossing level is also the minimum for $P_s$.

It should be observed that the fluctuation increases with the laser power $P$ (or $\langle n \rangle_1$). See Eq. (21). A higher fluctuation induces more errors for the attacker; fluctuation also increases with the gain $G$; $P$ and $G$ are easy parameters to control. Figure 26 shows $P_e$ as a function of $G, M$ and three values of the light power $P$.

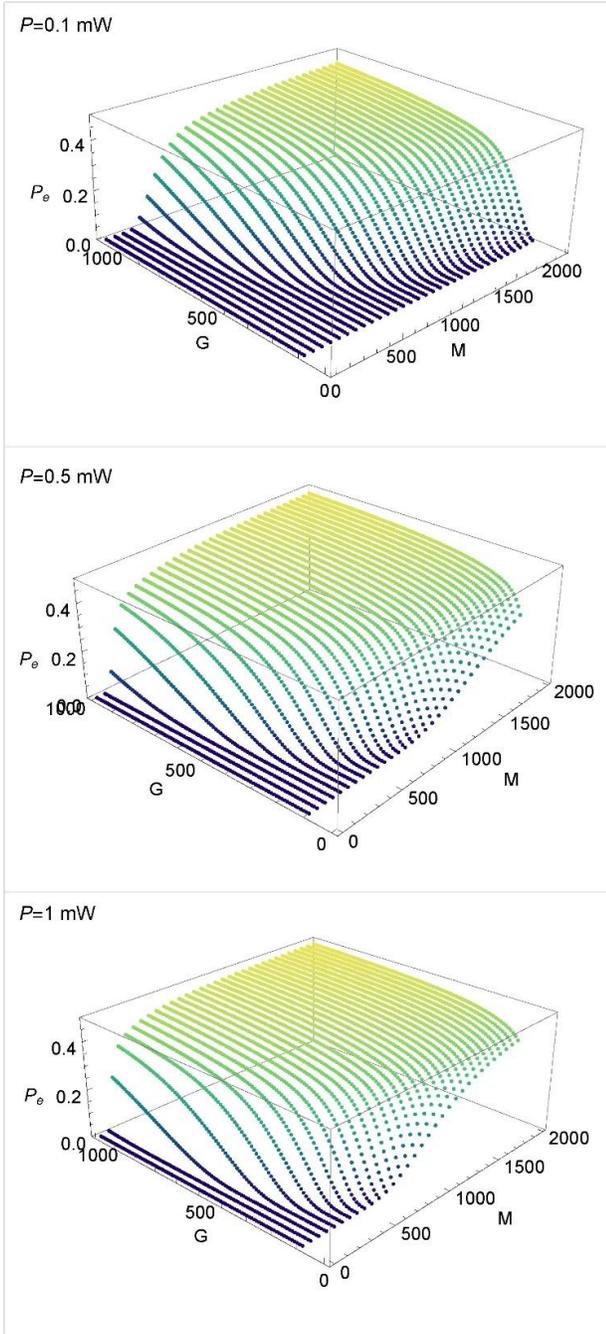

Fig. 26. $P_e$ as a function of $G, M$ and three light powers $P = 0.1 mW, 0.5 mW, 1 mW$.

### G. Probability of error at station B

One may assume that when subtracting the shared bases to extract the bits sent, an exceptionally large noise fluctuation has been recorded. With the encoded bit signal setting, the extracted signal close to the wrong encoded bit line causes a reading error by B. See Figure 10. The probability $P_{errB}$ for a so large fluctuation causing one error can be estimated from

$$P_{errB} = \frac{e^{-\frac{(jV_{max}/2M)^2}{2\sigma_V^2}}}{\sum_{j=0}^{M-1} e^{-\frac{(jV_{max}/2M)^2}{2\sigma_V^2}}} \tag{31}$$

Estimating the wrong line distant at $q \sim (3/4)M$ from the correct one, Figure 27 shows that $P_{errB} \to 0$. In other words, the error associated by the legitimate user hitting the wrong set of points is practically zero.

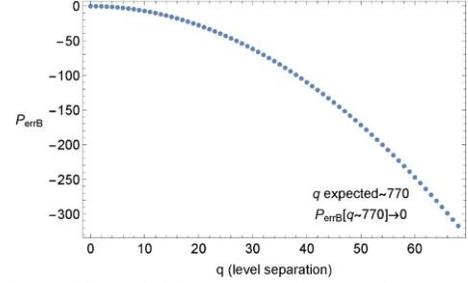

Fig. 27. 1 $log_{10}$ of B's probability of error $P_{errB}$ for a large noise fluctuation giving values around $(3/4)M$ for $M = 2^{10}$ and $P = 300 \mu W$. The probability is completely negligible.

### H. Mutual information and speed overhead

As discussed in Section V, for each signal sent there is a probability that the attacker succeeds in obtaining an amount of information $t_1$ about the bit. After $n$ bits sent, the total number of bits probabilistic leaked to the attacker will be $t = t_1 \times n$.

$t_1$ can be identified with the probability of success given by Eq. (29) and, therefore,

$$t = n \times \left(P_s - \frac{1}{2}\right). \tag{32}$$

Eq. (32) establishes that only bits with probabilities above $1/2$ could be leaked in a useful way to the attacker.

The PA protocol demands that this number $t$ of bits must be destroyed as well as an extra number $\lambda$ (=security parameter; see [27]. In other words, $t + \lambda$ bits are reduced from the set $n$. This has two main effects: one is that the distilled key $z$ will be shorter than the original length $a$; the other is that this reduction implies in a slowdown from the process of key generation in the PhRBG to the final encryption key $z$ at the output.

After signals are digitized, new probabilities should be calculated using the binary signals but analog results give a complete view of the security provided by the recorded quantum noise.

### I. Bits left after privacy amplification

The fraction $f$ of the bits left to be used compared to the initial input $n$ is

$$f = \frac{n - (t + \lambda)}{n} \tag{33}$$
$$= \frac{n - \left(n \times \left(P_s - \frac{1}{2}\right) + \lambda\right)}{n}.$$

Figure 28 illustrates the dependence of $f$ as a function of the number of levels $M$ and the security parameter $\lambda$.

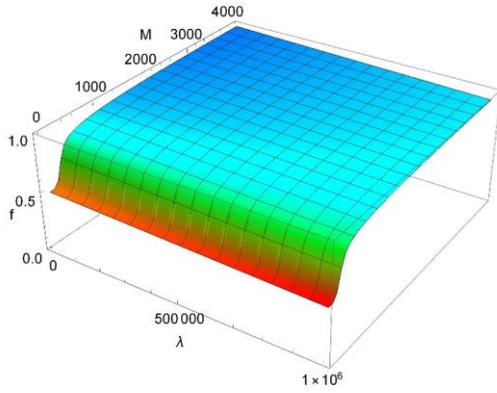

Fig. 28. Bit fraction $f$ left after privacy amplification as a function of the number of bases $M$ used and the number of bits used as the security parameter [27]. In this example $n \simeq 17 Gbit$.

*Figure 29*Fig. 29. - Bit fraction $f$ left for $n \cong 33 kbit$.

shows the bit fraction $f$ left for $n \cong 33 k$bit.

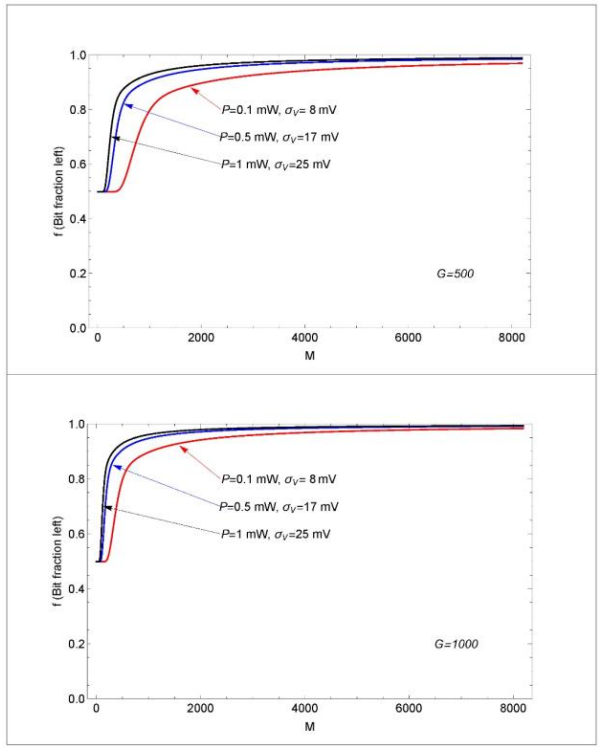

Fig. 29. - Bit fraction $f$ left for $n \cong 33 kbit$.

### J. Bits erased after privacy amplification

In other to emphasize the bit overhead produced by the PA protocol, Figure 30 shows the fraction erased from the initial number of bits, for the given conditions shown.

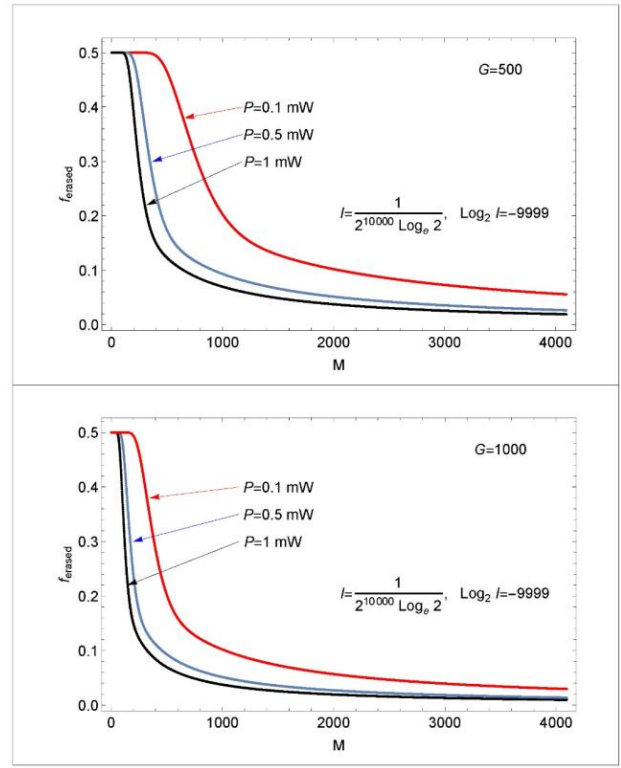

Fig. 30. Bit fraction erased $f_{erased}$ as a function of the number of levels $M$. In this example the security parameter is fixed and equal to $\lambda = 10000$ for $34 Gbit$ sent. The mutual information I is $2^{-9999}$. Notice that when the number of levels $M$ is too low, the number of bits that need to be erased is very high -a high overhead. However, for large $M$ a large number of bits remains. The higher number of useful bits demonstrates the need of an ADC with more bits (usually more expensive).

As the security parameter defines the mutual information value, a large λ results in high secrecy. At the same time to have a small fraction of rejected bits, the initial number $n = a + m \times a$ must be high. This suggests use of runs with long sequences of bits instead of short sequences in separate emissions. The size of these long sequences will be determined in practice by the existing FPGA memory capacity.

Summarizing, a practical perfect security is achieved for key distribution. This assures the perfect security for encryption with the distilled bits. The overhead for this process comes from the use of M-ry bases for each bit send (giving a $m = \log_2 M$ reduction in speed from the rate of key generation) plus the bit fraction erased from the original number of bits $n = a + m \times a \rightarrow r = n - (t + \lambda)$.

The overall gain in benefits is salient: the absence of distance limitation, use of any transmission channel including commercial networks, low cost, and the use of OTP without the need for couriers to refresh keys in a continuous process.

### K. Collision probability for the decentralized encryption

Section VII.A introduced decentralized encryption. The meaning is that a single Platform can generate the same set of keys for $N$ users and these users can encrypt messages among them not relying on the key source for each encryption done. This may be handy for members of a team that need to exchange messages among them, one-to-one or one-to-many, in a way that simplifies key management.

One may ask what happens if two users choose the same lines of encrypting keys (see Figure 13) to encrypt different messages. This is referred to as collision probabilities.

Define the collision probability [33] so that two among $N$ users choose a same line (one collision) within the matrix with $d (= \sqrt{K})$ lines:

$$P_{1Collision}(N, K) = 1 - \frac{d!}{(d-N)d^N} \quad (34)$$
$$\approx 1 - \left(1 - \frac{N}{2d}\right)^{N-1}.$$

Figure 31 shows $P_{1Collision}(N, K)$ for $N = 2$ to $20$ and $K = 10^7$ to $10^9$.

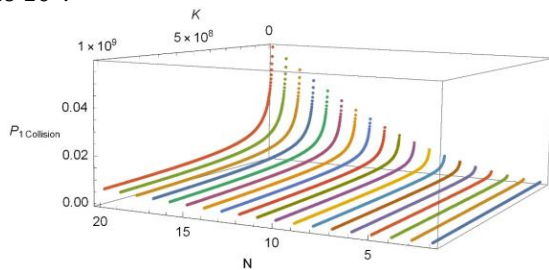

Fig. 31. Probability for one collision due to two users within $N$ users choosing the same line in the matrix representing the bits for decentralized encryption. In this example, $K$ goes from 10M bit to 1 Gbit.

Estimating that the order of magnitude for 20 collisions is $P_{1Collision}(N, K)^{20}$, Figure 32 shows the numerical results.

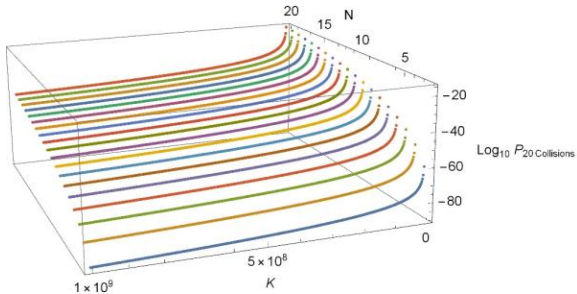

Fig. 32. Estimate of 20 collisions due to two users choosing same lines in the matrix representing the bits for decentralized encryption.

These numerical examples show that the decentralized encryption [28] provides a reliable and safe protocol that assures safe one-time-pad encryption/decryption among N users (assuming that the keys are safely stored and user access to storage is rigorously controlled).

After a number of times of uses of the original batch of keys, these should be refreshed. This number of usages could be equivalent to a single use of all keys.

*L. Comparisons of the KeyBITS generator and other PhRNG*

For encryption at large volumes, a Physical Random Bit Generator must be fast with a reliable randomness at the output. Using examples of encrypting modest size pictures, and at the rate of 100,000 pictures in one hour, close to 1 Terabyte of keys per hour are needed. There are not many competitors working with physical random number generators requiring fast rates.

Regarding Physical Random Bit Generators, among the existing technologies involving physical principles, those that exploit the quantum fluctuations of light are the most promising ones in terms of speed and true randomness (i.e., no algorithms to be explored by hackers).

To illustrate the advantages of our generator over existing commercial technologies, and even not-yet-developed systems described in published physics research , below is a short list of generators representing a broad spectrum of optical phenomena. Each is compared to KeyBITS:
1) Quantis (www.idquantique.com);
2) Quintessence (www.quintessencelabs.com)
3) Twin beams from Spontaneous Parametric Down Conversion (SPDC) (Applied Optics vol 48, No 9, 1774, 2009)
4) "Performance of Random Number Generators Using Noise-Based Superluminescent Diode and Chaos-Based Semiconductor Lasers" (IEEE Journal of Selected Topics in Quantum Electronics, Vol. 19, No. 4, July/August 2013)
5) Radioactive emitters: Apart from laboratory studies a recent advertisement announced a commercial random generator, EYL, based on radioactive decay.

For physical key generation an established competitor at low speeds (up to 16MHz) is the Quantis random number generator produced by ID-Quantique. For high speed, an example is the compact radioactive generator from EYL.

The Quantis generator uses a weak beam of laser light where the photons in a laser beam are split by a beam splitter and detected by two separate detectors signaling bits 0 and 1. It processes up to approximately 10Mbits/second and *cannot* be made faster due to a *physical limitation*: More than one photon can be generated in a laser beam within a sampling time. Therefore, events where simultaneous photons reach both detectors have to be discarded. If the intensity of the laser is increased to improve the system's bit generation, the equipment speed becomes useless due to this photon simultaneity in both detectors. This is an unsolvable limitation – not present in our system.

Quintessence splits a light beam, sending light to two detectors. In principle, it is not subject to a speed limitation because it works with multi-photons. The two detectors work with subtraction electronics to extract the random fluctuations of light. The limitation is that it demands a strict splitting balance of the light intensity that requires continuous management; otherwise the signal subtraction from both detectors becomes inefficient, and the noise fluctuations are covered by the intensity signals. Our system uses a single detector, no interferometry, and simpler electronics and therefore presents a more elegant, easily managed, and robust system.

Photon pairs generated by Spontaneous Parametric Down Conversion (SPDC) are superposed on a beam splitter, producing two-photon interferences. Single photon outcomes detected by four single photon detectors guarantee bit independence with no bias. However, the physical apparatus needed for this process cannot be miniaturized, it is costly and extremely slow for bit generation (below1Mbits/second). Therefore, it lacks commercial potential.

A research group at Saitama University proposed an 8.3 Gbits/second random number generator based on optical sources. Their design is experimental and relies on expensive lab instruments, such as a digital phosphor oscilloscope. Other research groups have proposed similar experimental methods, but are either based on specialized lab instruments or have a significantly lower generation rate.

Radioactive bit generators can be made compact but have an intrinsic speed limitation because increases in the rate of photon emissions rely on increasing the radiation level. Increases in radiation levels present health and safety risks and subjects the apparatus to regulatory constraints. The EYL generator works with clicks originated from radioactive processes at low radiation levels. For reasonable rates the amount of radioactivity can be kept low, but to increase the random bit output, the level of radioactivity must increase; an undesirable feature.

Figure 33 shows the developed KeyBITS PhRBG.

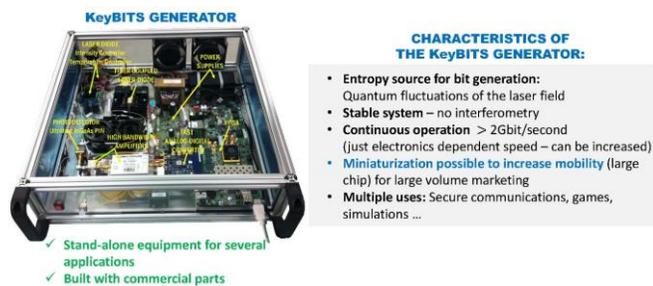

Fig. 33. PhRBG and properties

The KeyBITS generator can be miniaturized where needed to service an increasingly mobile workforce and operation. Its speed can be increased only depending on electronic advances, and no radioactivity is involved. Its simplicity, using only one detector and with no need for interferometric setups or subtle light balances have yielded a device with more desirable and utile characteristics than other PhRBG. Figure 34 compares the mentioned generators.

| Company or product | NIST tests Short sequences | NIST tests Long sequences | Large Bandwidth (fast speed) | Single detector: Simplicity + no need for balance | No radioactivity |
|---|---|---|---|---|---|
| ID Quantique | ✓ | ✗ | ✗ | ✗ | ✓ |
| Photon pairs | ✓ | ✓ | ✓ | ✗ | ✓ |
| EYL | ✓ | ✓ | ✓ | ✗ | ✗ |
| Quintessence | ✓ | ✓ | ✓ | ✗ | ✓ |
| **KeyBITS** | ✓ | ✓ | ✓ | ✓ | ✓ |

Fig. 34. Comparisons among some Physical Random Number Generators

It may be interesting to see that sequences of bits from the KeyBITS generator have been used as a standard to compare distinct encryption protocols [34].

*M. Completion of the KeyBITS Platform*

Currently, the keyBITS Platform and the software are being fully tested. The effect of the optical noise on the *M*-ry coded signals have been demonstrated in several papers; for example [3] to [25]. It is suggested that a testing network be established and that testing for security from attacks on the in-transit communication be done to demonstrate measurable results provided by these platforms. Results should be compared to those obtained from QKD experimental networks, such as the one controlled by Batelle Memorial Institute/Ohio (Columbus-Dublin network).

XI. ACKNOWLEDGMENT

Implementation of the prototype of the novel Physical Key Bit Generator (PhRBG) was made possible thanks to the support of the Brazilian grant 0276/12 from the Ministério da Ciência, Tecnologia e Inovação (MCTI), Financiadora de Estudos e Projetos (FINEP) and Fundação de Desenvolvimento da Pesquisa (FUNDEP), under a program originated from Comando do Exército (DCT)-RENASIC (Rede Nacional de Excelência em Segurança da Informação e Criptografia) coordinated by Colonel A. M. B. Monclaro) under Generals Eduardo Wolsky (CDS) and Paulo S. M. de Carvalho (CDCiber). Collaboration of Universidade Federal de Minas Gerais (UFMG) with Profs. Wagner N. Rodrigues, Roberto A. Nogueira, Júlio C. D. de Melo, Jeroen van de Graaf, Gilberto M. Ribeiro, and their students, was essential. Collaboration with Drs. Luiz F. Etrusco and Antonio O. Fernandes, from Invent Vision, is appreciated. Dr. Diana R. Carl and Prof. Oscar N. Mesquita are acknowledged for the critical reading of the manuscript and valuable suggestions.

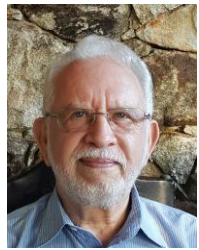

**Geraldo A. Barbosa** - PhD (Physics) / University of Southern California, Los Angeles / US, 1974. Areas of work: Quantum Optics and Condensed Matter (Theory and Experiment), Physical Cryptography.

Full Professor, Universidade Federal de Minas Gerais / MG / Brazil (up to 1995) / Northwestern University (2000-2012).

President – KeyBITS Encryption Technologies LLC/MD-USA.